\documentclass[twocolumn,aps,showpacs,prb,tightenlines,amsmath,amssymb,superscriptaddress]{revtex4}
\usepackage{graphicx}
\usepackage{bm}
\usepackage{dcolumn}

\usepackage{bm}

\usepackage{color}

\begin{document}

\title{Spin relaxation in multiple (110) quantum wells}
\author{M.M. Glazov}
\email{glazov@coherent.ioffe.ru}
\author{M.A. Semina} 
\affiliation{Ioffe Physical-Technical Institute, Russian Academy of Sciences, 194021 St. Petersburg, Russia}
\author{E.Ya. Sherman}
\affiliation{Department of Physical Chemistry, The University of the Basque Country, 48080 Bilbao, Spain} 
\affiliation{IKERBASQUE  Basque Foundation for Science, Alameda Urquijo 36-5, 48011, Bilbao, Bizkaia, Spain}

\date{\today}

\begin{abstract}
We consider theoretically the relaxation of electron spin component parallel to the growth
direction in multiple (110) GaAs quantum wells. The sources of spin
relaxation are the random Rashba spin-orbit coupling due to the electric field of donors 
and spin-flip collisions of electrons from different quantum wells. We show that the
screening of the Coulomb forces at low temperatures leads to a very strong
enhancement of the spin relaxation time. In a degenerate electron gas 
the Pauli blocking suppresses the electron-electron collisions,
and the leading spin relaxation mechanism comes from the field of donors. If the
electron gas is nondegenerate the electron-electron collisions and scattering by
the ionized donors give similar contributions to the relaxation rate.
\end{abstract}


\maketitle

\section{Introduction}

The ability to produce semiconductor systems with sufficiently long spin
relaxation time and spin diffusion length is expected to lead to a design of
reliable spin transport devices.\cite{Zutic04,Fabian07,Dyakonov08} Also, 
the problem of theoretical description of systems with a very long relaxation
time and understanding the physics of the limits for the spin relaxation rate
is of a fundamental interest. The major ingredient required for the spin
relaxation is spin-orbit coupling, which usually has the form of the wavevector 
dependent effective magnetic field. In quantum well structures there are two contributions to this field, namely,
linear in the in-plane momentum Rashba \cite{Rashba84} and Dresselhaus \cite{Dyakonov86} terms. One of
the fundamental aspects in this understanding is that the very low
relaxation rates for some spin directions are related to the specific symmetry of the
spin-orbit coupling Hamiltonian.\cite{Averkiev99,Bernevig06,Koralek09,Tokatly09}

One of the hopes at a possible realization of a very long spin relaxation
time system is related to the (110) GaAs quantum wells.
In these structures the form of the bulk inversion asymmetry (Dresselhaus) spin-orbit coupling 
prevents relaxation of the spin component parallel to the growth axis [110].  
Here the effective field is always collinear with this axis, \cite{Dyakonov86} and, therefore, the
spin, oriented along it, experiences no torque.
However, the experiment clearly demonstrates 
finite spin coherence lifetime \cite{Dohrmann,Oestreich,Belkov,Muller08}
causing the discussion of its origin.
A possible explanation can be related to the effects of intersubband scattering of 
electrons and intersubband spin-orbit coupling,\cite{Dohrmann,Wu2} which
become efficient at relatively high temperatures and depend strongly on the 
disorder in the quantum well. Calculations 
show the dephasing times {being} on the order of ten nanosecond at  $T=100$ K.\cite{Wu2}  With the temperature
decrease, the relaxation time rapidly increases, and for the
values of disorder, corresponding to the measured mobility, exceeds
the observed one by more than two orders of magnitude.\cite{Wu2} 
Therefore, other causes for the experimentally observed dephasing rates should be looked for.  

The possible mechanism can be related to the intrinsic disorder in the
Rashba coupling. If the carriers in the well come from 
the donors, the electric field of the ionized donors causes a random
Rashba field.\cite{Sherman03,Glazov05} The corresponding effective magnetic 
field is directed in the structure plane,
and, therefore, can influence the growth-direction component of the spin.
Its randomness causes random spin precession and leads to the finite spin
relaxation rate. If the Coulomb field
of the donors can be reduced, the relaxation time can be made longer. This can be
achieved by using multiple quantum wells, bringing about two new physical
effects not considered in previous works to the best of our knowledge. First, Coulomb forces can be strongly reduced by the screening due
to the presence of conduction electrons. Second effect in these structures
is the spin-flip scattering of electrons from different
quantum wells. In these collisions spin of one electron remains constant
while the spin of the other flips. This non-conservation arises due to the
spin-orbit coupling effect in the scattering process. It is very similar to the
spin-flip scattering by charged donors: screened electric field of electron
in one well induces a Rashba field acting on the electron spin in the other well. 
We focus on these two spin relaxation processes, inherent for 
multiple quantum wells and limiting the spin relaxation time
there.

We demonstrate here that in symmetric multiple quantum well structures grown along the 
$[110]$ axis, extremely long spin lifetimes, up to $\sim 100$~ns can be achieved at low temperatures.
These times exceed significantly the experimental values reported in the literature.
We discuss the origin of such a discrepancy and find that
even a very weak asymmetry of the quantum well structure can bring the 
spin lifetimes to the experimentally observed values.

This paper is organized as follows. In the second Section we introduce the
quantum well structure of interest and study the role of the screening
due to the presence of several quantum wells filled with mobile
electrons. In the third Section we study the spin relaxation 
due to the random Rashba field of screened donors as a function of temperature. 
In the fourth Section we investigate the
temperature-dependent effect of collisions of electrons from different quantum
wells on the spin relaxation and compare the spin relaxation rates due
to both effects. In Conclusions we summarize the results and discuss
possible experimental observations of the effects considered in this paper. 

\section{Quantum well structure and the Coulomb screening}

In Fig. 1(a) we present a scheme of a typical multiple quantum well structure:
relatively thin conducting layers separated by wide barriers. 
Such a structure is similar to the one studied in Ref.~[\onlinecite{Muller08}]. To reduce the
structure asymmetry-induced Rashba field, each quantum well is located in a
macroscopically symmetric environment. The electrons in the conducting
layers come from ionized donors located between them. A typical building
element of a multiple quantum well is a three-well structure
presented in Fig. 1. The central and the side quantum wells are separated by
the distance $L,$ typically on the order of 50-100 nm. The donor layers are
inserted symmetrically at the distance $z_{\rm ext}$ to prevent the appearance
of the macroscopic Rashba field in the central well{, so that} the structure in
Fig.~1 maintains macroscopic symmetry with respect to the $z\longleftrightarrow -z$
reflection. The central and the side wells are filled with conduction
electrons. The screening due to the redistribution of these electrons
strongly reduces the electric field of the donor ions in the wells. We
assume that the quantum wells are sufficiently narrow 
and can be considered as zero-width conducting layers.\cite{Ando,Kovalev}

\begin{figure}
\includegraphics[width= 0.45\textwidth]{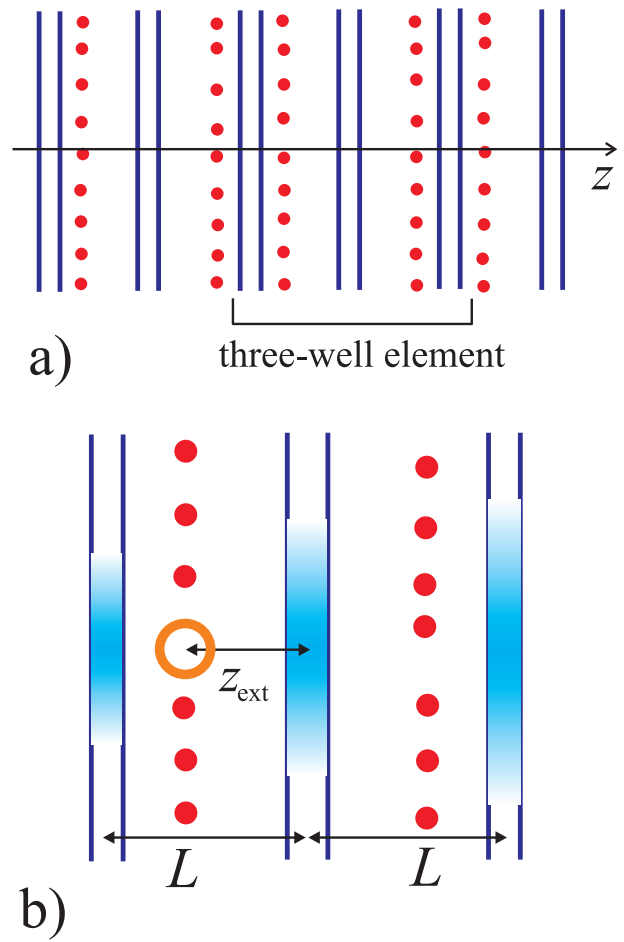}
\caption{(Color online).  (a) Multiple quantum well. Dopant layers are shown as filled circles. 
(b) Three-layer building element. The interwell distance is $L$, the distance between
the central well and donor layer is $\left|z_{\rm ext}\right|$. The selected
ion is presented by an empty circle. The nonuniformly filled rectangles 
sketch the electron density distributions induced by the selected ion 
in each quantum well.}
\label{structure}
\end{figure}

The fluctuating Rashba field caused by the ionized donors and the electrons
in the side wells is proportional to the $z$-component of the electric field
caused by the charges. As a result, it depends strongly on how this field
is screened. Electrons are free to move in the quantum well plane but cannot move along the growth direction, 
making screening of the $z-$axis and lateral field components drastically different. 
In a narrow single quantum well the $z$ component of the donor-induced field 
is not screened at all. In multiple quantum wells this component is reduced 
due to the redistribution of the electron density in the side wells.
In order to evaluate the screening, we find the static potential $U\left( {\bm r}\right) $ by solving the Poisson equation:
\begin{equation}
\varkappa \Delta U\left( {\bm r}\right) =-4\pi \left[ {\varrho} _{\rm ext}\left( 
{\bm r}\right) +{\varrho} _{\rm ind}\left( {\bm r}\right) \right],
\label{Poisson}
\end{equation}
where $\varkappa $ is the background dielectric constant, and $\varrho_{\rm ext}\left( {\bm r}\right) $ is the external charge density. The
corresponding induced charge density ${\varrho}_{\rm ind}\left( {\bm r}\right) $
at the layers numerated by index $l$ is given by: 
\begin{equation}
{\varrho}_{\rm ind}\left( {\bm r}\right) =-\frac{\varkappa }{2\pi }
\sum_{l}q_{s}^{[l]}U\left( {\bm r}\right) \delta \left( z-z_{l}\right)
 ,
\end{equation}
where summation is taken over all the layers $l$ in the system. The
concentration- and temperature-dependent two-dimensional Thomas-Fermi
screening wavevector $q_{s}^{[l]}$ can vary from well to well and is given
by: 
\begin{equation}
q_{s}^{{[l]}}=\frac{2}{a_{B}}\frac{1}{1+\exp \left( -\mu^{{[l]}} /T\right) }.
\label{qs}
\end{equation}
Here $a_{B}$ is the effective Bohr radius, corresponding to the electron
effective mass and the background dielectric constant, and $\mu^{{[l]}} $ is the
chemical potential of the electron gas in a given well. Due to the translational invariance
of the system in the lateral direction ${\bm \rho}$, it is convenient to use the
Fourier-transformed potential $U\left( q,z\right) $ corresponding to the
in-plane wavevector $q.$ For an external point charge ${\varrho}_{\rm ext}\left( 
{\bm r}\right) =e\delta ({\bm \rho_{\rm ext}})\delta \left( z-z_{\rm ext}\right) ,$ where $e$
is the elementary charge, selected in Fig. 1, the solution of the Poisson
equation (\ref{Poisson}) can be presented in the form: 
\begin{equation}
U\left( q,z\right) =\frac{2\pi e}{\varkappa }\left( \frac{1}{q}e^{-q\left|
z-z_{\rm ext}\right| }+\sum_{l}g_{l}(q)e^{-q\left| z-z_{l}\right| }\right) .
\end{equation}
Everywhere the normalization area is set to unity. Equation for the
components $g_{l}(q)$ can be obtained by a conventional method \cite{Ando,Kovalev} keeping only the singular parts in the second derivative 
of  $U\left( q,z\right)$ in Eq.~(\ref{Poisson}). As a result, we obtain the system of linear
equations: 
\begin{equation}
\sum_{l}\left( e^{-q\left| z_{m}-z_{l}\right| }+\frac{q}{q_{s}^{[m]}}\delta
_{ml}\right) g_{l}(q)=-\frac{1}{q}e^{-q\left| z_{m}-z_{\rm ext}\right| }.
\label{linear}
\end{equation}
>From this system one obtains the components $g_{l}(q)$ and, therefore, the field in the entire space. In order to calculate the $z$-component of the electric field in the central well, we take into account that the lateral redistribution of the electron density {there} does not cause net $z$-component of the electric field acting on electrons in this well. Therefore, this redistribution does not contribute in the screening of the  $z$-component. For this reason, we can  express the solution in terms of $g_l(q)$ with $l\ne 0$ only. 

In what follows we focus on the three-well structure, our results do not change qualitatively for a multiple well systems.
Then, for a point charge we obtain the Fourier-component of the electric
field in the layer $l=0$:
\begin{multline}
E_{z}\left( q,{z=0}\right) =-\mathrm{sign(}z_{\rm ext}\mathrm{)}\frac{2\pi e}{%
\varkappa }\times \\
\left\{ e^{-q\left| z_{\rm ext}\right| }+e^{-qL}q\left[ g_{-1}(q)%
\mathrm{-}g_{1}(q)\right] \right\} e^{i{\bm q}{\bm \rho}_{\rm ext}}.
\label{Ez}
\end{multline}
Below, for simplicity, we omit the central layer coordinate and for $%
{\bm \rho }_{\rm ext}=0$ obtain from Eqs.~(\ref{linear}), (\ref{Ez}): 
\begin{equation}
E_{z}(q)=-\mathrm{sign(}z_{\rm ext}\mathrm{)}\frac{2\pi e}{\varkappa }\frac{%
(q+q_{s})e^{-q\left| z_{\rm ext}\right| }-q_{s}e^{-2qL+q\left| z_{\rm ext}\right| }}{%
q+q_{s}\left( 1-e^{-2qL}\right) },
\end{equation}
where we assumed that the screening wave vectors are the same in side wells, $q_{s}^{[-1]}=q_{s}^{[1]}\equiv q_s$, 
otherwise the structure is asymmetric and one has to allow for the regular Rashba field as well. 
The field $E_{z}(q)$ has the following important features.
First, if the charge is close to the central quantum well, that is $\left|
z_{\rm ext}\right| \ll L,$ and $qL{\gtrsim} 1$, the induced field of the side layers
can be neglected, and, therefore, the screening does not play a significant
role. Second, if the external charge is close the side plane, that is $%
L-\left| z_{\rm ext}\right| \ll L,$ and the screening is strong enough, such as $%
\left( L-\left| z_{\rm ext}\right| \right) q_{s}\gg 1,$ the resulting field can
be to a good approximation described by a dipole formed by the external
charge and the opposite charge induced at the nearest side plane.

\section{Spin relaxation in the random Rashba field of donors}

Fluctuations in the donor density form in the central well a local Rashba
spin-orbit coupling proportional to their local field $E_{z}\left({\bm\rho}\right).$ 
These fluctuations cause a random precession of the spin,
and, as a result, spin relaxation. The spin-orbit Hamiltonian for the system
under study can be presented as the sum of the regular Dresselhaus and a
random Rashba terms as:
\begin{eqnarray}
H_{so} &=&H_{D}+H_{R}, \\
H_{D} =\frac{\hbar }{2}\left( {\bm \Omega }_{D}\cdot {\bm \sigma }%
\right) ,& &\quad H_{R}=\frac{\hbar }{2}\left( {\bm \Omega }_{R}\left({\bm \rho }\right)\cdot 
{\bm \sigma }\right) , \\
{\bm \Omega }_{D} &=&\frac{2}{\hbar }\alpha _{D}(0,0,k_{x}),  \nonumber\\
{\bm \Omega }_{R}\left({\bm \rho }\right) &=&\frac{2}{\hbar }\alpha
_{R}\left({\bm \rho }\right) (k_{y},-k_{x},0),
\label{Omegas}
\end{eqnarray}
where ${\bm k}$ is the in-plane momentum, and $\bm \sigma$ is the vector composed of the Pauli
matrices. Here $\alpha _{D}$ is the Dresselhaus coupling constant, dependent on
the quantum well width, and the axes are chosen as: 
$x\parallel [1\overline{1}0],$ $y\parallel [001],$ and $z\parallel [110]$. The precession rate due to
the Rashba coupling ${\bm \Omega }_{R}\left( {}{\bm \rho }\right) $ is
a random function of the lateral coordinate, providing the only contribution
leading to the relaxation of the $z$-component of averaged electron spin. We assume here the classical
spin-orbit potential, that can be done for $k\left| z_{\rm ext}\right| \gg 1,$
neglecting non-commutativity of the momentum and coordinate-dependent Rashba
field.\cite{Dugaev09} 
We  take the Rashba parameter being proportional to the local electric field: 
\begin{equation}
\alpha_{R}\left({\bm \rho }\right) =\xi eE_{z}\left( {\bm \rho }%
\right),
\label{alphafield}
\end{equation}
where $\xi$ is the structure-dependent constant.\cite{Harley,Winkler}

\begin{figure}
\includegraphics[width= 0.45\textwidth]{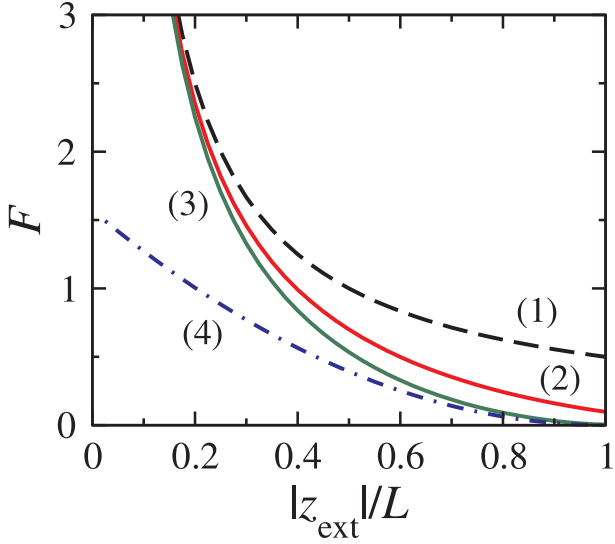}
\caption{(Color online).  Plot of dimensionless correlator $F$ as a function of donor layer 
position $\left|z_{\rm ext}\right|/L$ for different system parameters. Dashed line (1) $F=L/(2\left|z_{\rm ext}\right|)$ corresponds to the
absence of the screening, lines (2) and (3) correspond to $q_sL=1.0$, and $q_sL=10$, respectively, and dashed-dot line
(4) presents the field of the dipole formed by the point donor charge and the induced charge in the side well, 
which corresponds to $q_sL\to \infty$. 
{At $\left|z_{\rm ext}\right|/L\ll 1$ lines (1),(2), and (3) converge, demonstrating that
the screening becomes unimportant in this regime.}}
\label{impurities_1}
\end{figure}

We begin with the reminder of the spin relaxation mechanism due to the
donors producing random electric field. The Fermi golden rule shows that
the relaxation rate for the $z-$component of the spin $\Gamma _{d}^{[s]}(k)$
for electron with wavevector $k$ can be expressed as the integral of the
precession rates correlator\cite{Semenov} in the random Rashba field: 
\begin{equation}
\Gamma _{d}^{[s]}(k)=\int_{0}^{\infty }\left\langle {\bm \Omega }%
_{R}\left( t\right) {\bm \Omega }_{R}\left( 0\right) \right\rangle dt.
\label{integral}
\end{equation}
For the straightforward ballistic motion of electron with ${\bm \rho }=%
{\bm v}t$, which can be used if the electron free path is much larger
than $\left| z_{\rm ext}\right|$, the relaxation rate can be expressed using Eqs.~\eqref{Omegas}, \eqref{alphafield} as: 
\begin{eqnarray}
\Gamma _{d}^{[s]}(k) &=&\frac{4}{\hbar^{2}}\xi^{2}e^{2}k^{2}\,\int \left\langle
EE\right\rangle _{q}\frac{d^{2}q}{\left( 2\pi\right) ^{2}}\int_{0}^{\infty
}e^{i{\bm q}{\bm v}t}dt \nonumber \\
&=&\frac{2}{\pi{\hbar^2}}\xi ^{2}e^{2}\frac{k^{2}}{v}\int_{0}^{\infty }\left\langle
EE\right\rangle _{q}dq,
\end{eqnarray}
where the Fourier-transform of the correlator $\,\left\langle E_{z}\left( 
{\bm \rho }\right) E_{z}\left( {\bm 0}\right) \right\rangle $ entering
Eq. (\ref{integral}) is written as: 
\begin{equation}
\,\left\langle E_{z}\left( {\bm \rho }\right) E_{z}\left( {\bm 0}%
\right) \right\rangle \equiv \int \left\langle EE\right\rangle _{q}
e^{i{\bm q}{\bm\rho}}\frac{d^{2}q}{\left( 2\pi \right) ^{2}}.
\end{equation}
The resulting spin relaxation rate for electron with wavevector $k$ can be
presented as: 
\begin{equation}
\Gamma _{d}^{[s]}(k)=8\pi \frac{\xi ^{2}km}{\hbar ^{3}}\frac{e^{4}}{%
\varkappa ^{2}L}n_{d}{F},
\label{rate}
\end{equation}
with $n_{d}$ being the concentration of donors per layer, and the
dimensionless integrated correlator: 
\begin{equation}
{F}=L\int_{0}^{\infty }f_{q}dq,
\end{equation}
with 
\begin{equation}
f_{q}=\left[ \frac{(q+q_{s})e^{-q\left| z_{\rm ext}\right|
}-q_{s}e^{-2qL+q\left| z_{\rm ext}\right| }}{q+q_{s}(1-e^{-2qL})}\right] ^{2}.
\end{equation}
Here we assume the white noise 
$\left\langle n({\bm \rho}_{1})n({\bm \rho}_{2})\right\rangle$ $=n_{d} \delta \left({\bm\rho}_{1}-{\bm \rho}_{2}\right)$
random distribution of the ion density. 

It is clearly seen in Eq.~\eqref{rate} that the spin relaxation due to 
the fluctuating Rashba field occurs even at straightforward electron motion and does not require, in general, the momentum scattering. The relaxation rate can be estimated as a product of squared fluctuation of the random spin precession frequency, $\Omega_R^2$, and the passage time of the correlated domain of 
spin-orbit coupling,\cite{Sherman03, Glazov05} making spin and 
momentum relaxation processes independent, by contrast to the conventional Elliott-Yafet mechanism.~\cite{OptOr,agw}

Eq.~\eqref{rate} shows that the spin flip rate, being 
quadratic in the strength of the Coulomb interaction, 
is proportional to the donor density, the wave vector $k$ 
and the form-factor $F$. The dependence of numerically
integrated correlator ${F}$ on the distance between the donor layer
and the central well is presented in Fig. 2 for different values of $q_{s}.$
As one can
see in the Figure, the relaxation rate rapidly decreases and almost vanishes
when the donors become close to the side well, where the screening
becomes efficient. The role of the screening can be seen from a comparison
of the dashed (no screening) and solid (with the screening) lines: at $\left|z_{\rm ext}\right|>L/2$ all solid
lines are located well below the dashed one.

\begin{figure}
\includegraphics[width= 0.45\textwidth]{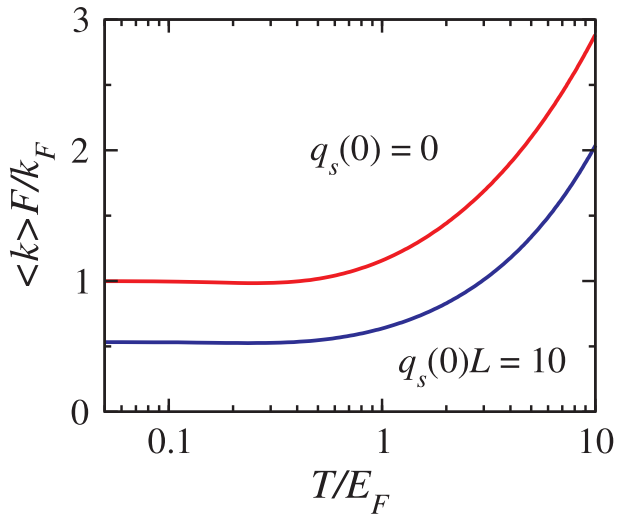}
\caption{(Color online).   The temperature dependence of $\left( \left\langle k\right\rangle /k_{F}\right)F$ 
for different screening parameters (their values at $T=0$ are marked near the plots). The donors are 
located at $|z_{\rm ext}|=L/2$.}
\label{impurities_2}
\end{figure}

The temperature dependence of the relaxation rate for the Fermi gas is given
by the formula:\cite{Dyakonov} 
\begin{equation}
\frac{1}{\tau _{d}^{[s]}}\equiv 2\left\langle \Gamma
_{d}^{[s]}(k)\right\rangle =2\frac{\sum_{\bm k}\Gamma _{d}^{[s]}(k)\left(
\partial f_{\bm k}/\partial E\right) }{\sum_{\bm k}\left( \partial f_{\bm %
k}/\partial E\right) },
\label{average}
\end{equation}
where $f_{\bm k}$ is the Fermi-Dirac distribution, and factor 2 in front
corresponds to the sum of contributions of two donor layers in the
relaxation rate. The result can be presented in the form: 
\begin{equation}
\frac{1}{\tau _{d}^{[s]}}=16\pi \frac{\xi ^{2}k_{F}m}{\hbar ^{3}}\frac{e^{4}%
}{\varkappa ^{2}L}n_{d} \frac{\left\langle k\right\rangle }{k_{F}}{F},
\end{equation}
where the angular brackets stand for the averaging similar to Eq.~(\ref{average}). There
are two main causes for the $T-$dependence of $1/\tau _{d}^{[s]}$. First, $%
\Gamma _{d}^{[s]}(k)$ depends linearly on the electron momentum. This
dependence will give a factor in the spin relaxation rate proportional to
the expectation value $\left\langle k\right\rangle$. In addition, the
screening wavevector $q_{s}$ decreases with the temperature, as given by Eq.~(\ref{qs}). The
calculated dependence of $\left( \left\langle k\right\rangle /k_{F}\right) 
{F}$ on $T$ is presented in Fig.~\ref{impurities_2}. 
At low temperatures $T\ll E_{F}$ spin relaxation rate is very weakly
temperature-dependent and is given by $2\Gamma _{d}^{[s]}(k)$ 
from Eq.(\ref{rate}) with $k=k_{F},$ and
zero-temperature $q_{s}=2/a_{B}.$ At high temperatures $T\gg E_{F}$ where
the screening becomes unimportant, {$q_{s}$ is small,} we obtain: $f_{q}=e^{-2q\left|
z_{\rm ext}\right| }$ and ${F}=L/(2\left| z_{\rm ext}\right| ).$ Therefore,
the mean spin relaxation rate in this regime is: 
\begin{equation}
\frac{1}{\tau _{d}^{[s]}}=4\pi ^{3/2}\frac{\xi ^{2}k_{T}m}{\hbar ^{3}}\frac{%
e^{4}}{\varkappa ^{2}\left| z_{\rm ext}\right| }n_{d} ,
\end{equation}
where $k_{T}=\sqrt{2mT}/\hbar$, and $1/\tau _{d}^{[s]}$ increases as $%
T^{1/2}.$ {The temperature at which the screening effects become weaker 
can be estimated with condition $q_{s}L/2=1$ for typical $z_{\rm ext}=L/2$. Taking into
account that usually $q_{s}(0)L$ is on the order of ten, the role of the screening is sufficiently reduced
only in a nondegenerate gas at $T\gg E_F$; since in this regime $q_{s}\sim q_{s}(0)\times(E_F/T)$, 
temperatures $T>10E_F$ are required.}

\section{Electron-electron collisions}

\begin{figure}
\includegraphics[width= 0.45\textwidth]{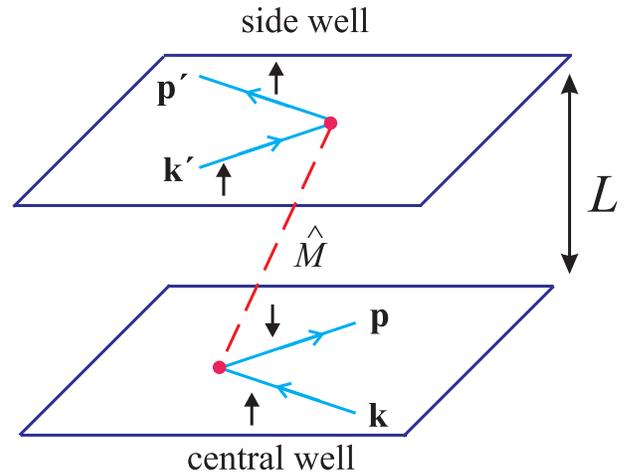}
\caption{(Color online).  Spin-flip electron-electron scattering process. Dashed line marked with $\hat{M}$ corresponds
to the scattering matrix element in Eq.\eqref{Coulomb1}.}
\label{ee_process}
\end{figure}

\smallskip Another source of spin relaxation can be the spin-flip
electron-electron collisions shown in Fig. \ref{ee_process}. Physically, this process can
be interpreted as the random spin precession of a given electron in the
fluctuating spin-orbit field caused by the other electron.

The corresponding matrix element of the spin-flip electron-electron
interaction can be written following Refs.~[\onlinecite{Glazov3,Glazov4}]:
\begin{multline}
\hat{M}=\delta _{\bm k+\bm k^{\prime },\bm p+\bm p^{\prime }}\frac{\xi {e}
E_{z}(q)}{2} \times \\ \left\{ [\hat{\bm \sigma }^{(1)}\times (\bm p+\bm k)]_{z}-[\hat{%
\bm \sigma }^{(2)}\times (\bm p^{\prime }+\bm k^{\prime })]_{z}\right\} ,
\label{Coulomb1}
\end{multline}
where $\bm k,\bm p$ are the initial and final wave vectors of the first
electron and $\bm k^{\prime },\bm p^{\prime }$ are the initial and the final
wave vectors of the second electron, $q=|\bm k-\bm p|$, $\hat{\bm \sigma }^{(1)}$, $\hat{\bm %
\sigma }^{(2)}$ are the spin-operators of the first and second electron,
respectively.  The Fourier transform of the $z$-component of the electric
field $E_{z}(q)$ is described by Eq.~(\ref{Ez}) with the interwell distance $L$ instead
of the distance to the donor layer $\left| z_{\rm ext}\right|.$
Hence, the electric field of electrons is weaker than the field of the ions and it is
considerably reduced by the screening. Note, that the collisions within the
same well do not lead to the $z$ spin component relaxation because $E_{z}=0$
in this case, and bulk inversion asymmetry contribution to $\hat{M}$ is
proportional to $\hat{\sigma}_{z}^{(1)}$, $\hat{\sigma}_{z}^{(2)}$ in $[110]$%
-grown quantum wells.\cite{Glazov4}
The calculation of the spin relaxation rate $1/\tau _{ee}^{[s]}$ is carried
out by using the kinetic equation for the spin density matrix. The
electron-electron collision integral for the scattering of electrons from
different wells is expressed via the properly antisymmetrized matrix element 
$\hat{M}$ using Keldysh technique.\cite{Glazov,Glazov1} The
resulting expression for the spin relaxation rate can be recast as 
\begin{equation}
\frac{1}{\tau _{ee}^{[s]}}=\frac{16}{\pi }\frac{me^{4}}{\hbar^{3}\varkappa
^{2}}k_{F}^{4}\xi ^{2}R,  \label{ee:gen}
\end{equation}
where the dimensionless factor $R$ depends on the temperature, electron
density, screening wavevector, and interwell distance. 

\begin{figure}
\includegraphics[width= 0.45\textwidth]{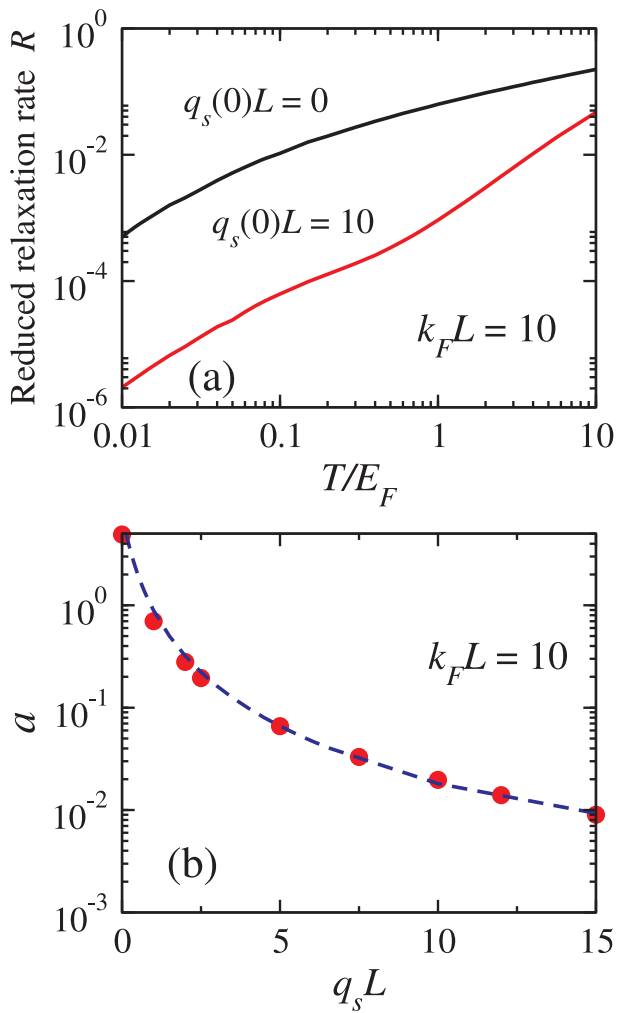}
\caption{(Color online).  (a) Temperature dependence of the reduced spin relaxation rate $R$ due to the electron-electron
scattering for different screening parameters marked near the plots. (b) Calculated values of the coefficient
$a$ as a function of screening (circles). Dashed line is $8.0/(1+2q_{s}L)^2$. $k_{F}L=10$ for both panels. }
\label{ee_results}
\end{figure}

At low temperatures, $T\ll E_{F}$, the electron gas is degenerate
and the electron-electron scattering is strongly suppressed by the Pauli
exclusion principle. The temperature dependence of the constant $R$ is shown
in Fig. \ref{ee_results}(a). There are two regimes of electron-electron scattering for
degenerate gas, different by the possible momentum transfer at the
scattering process. The first regime, when the momentum transfer is much
smaller than $1/L$ is realized at very low temperatures $T\ll E_{F}/(k_{F}L)$, and a typical Fermi-liquid behavior: 
\begin{equation}
R=a\times \left( \frac{T}{E_{F}}\right) ^{2}.
\end{equation}
is restored. In this regime one can take the $q\rightarrow 0$ limit in Eq.~(\ref{Ez}), and
obtain the coefficient $a$ in the form $c/(1+2q_{s}L)^{2}$ with the constant 
$c\approx 8.0$ as obtained by the fitting of the numerical results to this
simple analytic expression [see Fig.~\ref{ee_results}(b)]. At higher temperatures $%
E_{F}/(k_{F}L)\ll T\ll E_{F}$ the wave vector transferred in the process of
the Coulomb collision cannot exceed $\sim 1/L$ which reduces the phase
space for the final state. Therefore $R$ becomes linear in the temperature, 
as can be seen in Fig.~\ref{ee_results}(a), where the slope $d\ln R/d\ln T$ 
gradually changes from 2 to 1 with the temperature increase. 

At the temperatures close to the Fermi energy the transition to the
non-degenerate gas occurs and the screening reduces, leading to the
enhancement of the electron-electron scattering. At high temperatures $T\gg
E_{F}$ where the screening vanishes with $q_{s}=0,$ the calculations can be
carried out analytically. The spin relaxation rate in this regime is given
by 
\begin{equation}
\frac{1}{\tau _{ee}^{[s]}}=\frac{6\pi ^{3/2}}{\sqrt{2}}\frac{\xi ^{2}k_{T}m}{%
\hbar ^{3}}\frac{e^{4}}{\varkappa ^{2}L}n_{el},
\end{equation}
where $n_{el}$ is the concentration of electrons per single side well, where 
we assume for simplicity equal electron filling in all wells. This
situation is very similar to the scattering by charged donors, and $1/\tau
_{ee}^{[s]}$ increases as $T^{1/2}$. The ratio of the relaxation rates due
to donors and electron-electron collisions tends, therefore, to a
system-dependent constant: 
\begin{equation}
\frac{\tau _{ee}^{[s]}}{\tau _{d}^{[s]}}=\frac{2\sqrt{2}}{3}\frac{L}{|z_{\rm ext}|}%
\frac{n_{d} }{n_{el}},
\end{equation}
and is very close to $1$ when the electron and ion densities are equal and $L=|z_{\rm ext}|$. 
Therefore, the donors and electron-electron collisions give
contributions of the same order of magnitude in a nongenerate gas. 

As discussed above, at low temperatures, where the electron gas is
degenerate, one can expect a less efficient contribution of spin-flip
electron-scattering compared to the random Rashba fields of the donors.\cite{impurities}
To provide an example, if donors are close to the middle of the
interwell barriers, at $T=E_{F},$ typical contribution of electron-electron
collisions is two orders of magnitude weaker than that of the random Rashba
field of donors. With the decrease in the temperature, the relative effect
of collisions compared to the role of donors decreases. 

\section{Discussion and conclusions}

We investigated the electron spin relaxation in a symmetric three-layer GaAs (110) quantum well
structure, which can be considered as a building element for larger multiple quantum
well structures. Since the Dresselhaus spin-orbit {term} 
in these systems is proportional to $\sigma_{z}$, the $z$-component of the spin
demonstrates a very long lifetime. The spin relaxation can appear due to the random Rashba spin-orbit coupling 
resulting from the electric field of donors providing 
electrons for the quantum wells filling. The screening of the Coulomb interaction 
by the degenerate gas of conduction electrons strongly suppresses the field of the
ions and the corresponding random Rashba field. As a result, the spin relaxation time 
for an appropriately chosen geometry 
increases by at least an order of magnitude compared to a single symmetric quantum well. 

Another effect important in these structures is the
non spin conserving electron-electron collision, having the same physical
reason as the spin-flip scattering by charge donors, that is the local
Rashba field induced by the electric field of a charge located near a
quantum well. The contribution of electron-electron collisions is suppressed
by the Pauli principle at low temperatures, where the electron gas is
strongly degenerated, being much smaller than the contribution of the 
ions. At high temperatures, the scattering by the donors and by the electrons give
contributions to the spin relaxation rate of the same order of magnitude.

Here we concentrated on the spin relaxation in the symmetric central well.
Two other wells {can} be in a strongly asymmetric environment, and,
therefore, demonstrate a conventional relatively fast spin relaxation due to the
regular Rashba field. In the multiple quantum wells structures containing
many single layers, only edge wells are in the asymmetric environment. Therefore,
very long spin relaxation times can be achieved for the most of the electrons in the
system. 

These times can be estimated taking into account that for the low temperatures 
the main contribution comes from the ions. Eq.~(\ref{rate}) with the typical parameters
of GaAs-based quantum wells, and $\xi=5$ \AA$^2$ (Refs.~[\onlinecite{Harley,Winkler,Glazov3}])
yields the value of $\tau_s$ close to 400 nanosecond
at $|z_{\rm ext}|=L/2$. 
{A comparison of lines (1) and (3) in Fig.~\ref{impurities_1} shows that  
such a long time is considerably attributed to the screening; in a single well geometry 
with the same $|z_{\rm ext}|$ it would be smaller by approximately a factor of 2.} 
The experiment reveals the times an order of magnitude 
shorter.\cite{Muller08} 
The difference can be related to the nonprecise information about
the material and structure-dependent parameters and possible structure asymmetry resulting in a very weak
but regular Rashba field on the spatial scale comparable or larger than 
the electron free path. For the single-electron momentum relaxation time $\tau_p^{*}$ on the
order of 1 ps, it is sufficient to have a very small spin splitting at the Fermi level
on the order of 0.01 meV to ensure the times observed in the experiment.\cite{Muller08}
The contributions of the random and the regular Rashba mechanisms can be separated by studying the
temperature dependence of the relaxation rate at the temperatures $T\ll E_F$, where, however,
the single electron momentum relaxation time is already determined by the electron-electron collisions rather than
by the disorder.  In this case, the pure random Rashba field mechanism would result in a very weak
$T$-dependence, while the regular one would lead to a $1/T^2$ decrease.  

At these momentum relaxation times $\tau_p^{*}$, typical for $T$ on the order of 10-20 K, the electron free path is
larger than five interwell distances. Therefore, if the expectation value 
of the Rashba field is zero, at these temperatures electron spin indeed 
experiences only fluctuations of the Rashba field around its zero mean value. 
With the temperature increase, free path decreases and becomes comparable to
the interwell distance, being however still much smaller than the spin diffusion length.
At this temperature the electron gas demonstrates crossover to the conventional Dyakonov-Perel'
mechanism of spin relaxation with the effective averaged coupling $\left<\alpha_R^2({\bm \rho})\right>$ (Ref.~[\onlinecite{Wu1}]). A microscopic model of spin relaxation in the presence 
of nonvanishing Rashba coupling in [110] systems was developed  in Ref.[\onlinecite{Tarasenko09}]. 

We conclude from our analysis that by improving quality and design of multiple 
(110) quantum wells one can still achieve spin relaxation times an order of magnitude longer than the 
longest times observed so far.

\section{Acknowledgment}

EYS is grateful to the University of Basque Country UPV/EHU for support by
the Grant GIU07/40 and to the Ministry
of Science and Innovation of Spain for grant FIS2009-12773-C02-01. Financial support by RFBR, Federal program on support of 
leading scientific schools, President grant 
for young scientists and the ``Dynasty'' Foundation -- ICFPM is acknowledged.


\begin{thebibliography}{99}
\bibitem{impurities}  It is worth noting that in the case of the
ionized donor scattering the spin relaxation rate can be reduced to 
Eq.~\eqref{ee:gen} with the parameter $R=\pi\langle k/k_{F}\rangle {F}/(2k_{F}L)$.
\end{thebibliography}
\end{document}